\documentclass[12pt]{article}   
\usepackage{epsfig}
\usepackage{hangcaption}
\usepackage{array}

\newcommand{\dslash}{\partial\!\!\!/}

\begin{document}
\begin{flushright}
\end{flushright}

\vspace{1cm}
\begin{center}
\begin{large}
{\bf Higher Dimension Operators and Chiral Symmetry Breaking in
       Nambu--Jona-Lasinio Model}\\
\end{large}
\vskip 1.0in

Taekoon Lee%
\footnote{E-mail address: \tt tlee@ctp.snu.ac.kr}
 and Yongseok Oh%
\footnote{E-mail address: \tt yoh@mulli.snu.ac.kr}

        {\small {\it Center for Theoretical Physics, 
        Seoul National University,  Seoul 151-742, Korea}}
\end{center}
\date{}
\vskip 1cm
\begin{abstract}

We point out that adding higher dimension operators to the
Nambu--Jona-Lasinio model can stabilize chiral symmetry breaking and
resolve the issue raised by Kleinert and Van den Bossche that chiral
symmetry breaking cannot occur for $N_c \le N_c^{{\rm cr}}$ due to
large quantum fluctuations of the composite fields. 
\\[1cm]
PACS number(s): 11.30.Qc, 11.30.Rd, 12.39.Fe

\end{abstract}

\vskip 1cm
\def\thepage{SNUTP-99-046}
\thispagestyle{myheadings}
\newpage
\pagenumbering{arabic}
\addtocounter{page}{0}


The Nambu--Jona-Lasinio (NJL) model has been an important tool in
studying chiral symmetry breaking \cite{nl62,vw,klevansky92,hk94}.
The model is simple, comprised of four-fermion contact interactions,
and solvable in the large $N_c$ (number of colors) limit, showing chiral
symmetry breaking in a very clean manner \cite{klei}. 
Recently, however, Kleinert and Van den Bossche (KV) \cite{kb99} argued
that the strong quantum fluctuations of the  mesons around the
(classically) broken vacuum in the low energy effective Lagrangian of
the NJL model could restore chiral symmetry.
This chiral symmetry restoration occurs when the so-called
{\it stiffness\/}, which is the pion decay constant squared in the NJL
model, becomes smaller than a certain critical value.
This happens in the real world case $N_c=3$ and can cause a serious problem
for using the NJL model as a low energy effective description of the chiral 
symmetry breaking in QCD.

The four-fermion contact interactions  of the NJL model, which makes the
model nonrenormalizable, may be thought as  effective interactions arising
from a more fundamental one, for example from an exchange of massive 
intermediate particles in a renormalizable theory \cite{bee92,lkr96}. 
In this case, generally there would be higher dimension operators
(HDO's) in addition to the contact interactions.
These operators will be suppressed by the cutoff of the model, and at
tree level have small effects and can be ignored.
However, they cannot be ignored at loop level as long as momentum
cutoff is used in regularization.
This becomes clear when we recall that some of the loop diagrams in the
NJL model are quadratically divergent, and thus in momentum cutoff the
divergence could cancel the cutoff suppression of the HDO's, giving
unsuppressed contribution to the low energy effective Lagrangian.

The importance of HDO's in the NJL model was previously emphasized by
several authors \cite{aa93,hk}, while it was first discussed by Suzuki
\cite{suzuki90} in a specific form of dim$=\!\!8$ operators and by
Hasenfratz et al. \cite{hhjks91} in a more generalized form in the context
of top quark condensation model of electroweak symmetry breaking.
In this note we investigate the effects of HDO's on the chiral symmetry
breaking  and  point out in particular that adding proper HDO's to the NJL
model could suppress the quantum fluctuations of the mesons and prevent
chiral symmetry restoration.

We start with a two flavor NJL model:
\begin{equation}
{\cal L} = \bar{\psi} i\partial\!\!\!/ \psi + \frac{g}{2N_c\Lambda^2}
\left[ (\bar{\psi}\psi)^2 -(\bar{\psi}\gamma_5\tau_a\psi)^2\right],
\label{e1}
\end{equation}
where the massless quarks $\psi\equiv (u,d)^i$, $i=1,\cdots,N_c,$ carry
$N_c$ colors, $\tau_a$ are the Pauli matrices, $g$ is a dimensionless
coupling and $\Lambda$ is the cutoff.
Introducing auxiliary scalar ($\sigma$) and pseudoscalar ($\pi_a$) fields 
the Lagrangian can be written as
\begin{equation}
{\cal L} = \bar{\psi}( i\dslash - \sigma -i\gamma_5\tau_a\pi_a)\psi
- \frac{N_c\Lambda^2}{2g}(\sigma^2+\pi_a^2).
\label{e2}
\end{equation}
Integrating out the quark fields we obtain an $O(4)$ symmetric effective
Lagrangian for the mesons
\begin{equation}
{\cal L} = \frac{Z(\rho^2)}{2} \left[ (\partial_\mu\sigma)^2 +
(\partial_\mu\pi_a)^2 \right] - V(\rho^2) +
{\cal L}'(\sigma,\partial_\mu\sigma, \pi_a,\partial_\mu\pi_a),
\label{e3}
\end{equation}
where $\rho^2=\sigma^2+\pi_a^2$, while each term in ${\cal L}'$ has at
least four derivatives and inverse power suppressed in the cutoff
$\Lambda^2$, thus not of our interest.
The wave function normalization $Z(\rho^2)$ arises from Fig. 1 and is
given by
\begin{eqnarray}
Z(\rho^2) &=& \left.\frac{d}{dq_E^2} \Sigma(q_E^2) \right|_{q_E^2=0}
\nonumber \\
&=& \left.\frac{d}{dq_E^2} \left[8 N_c \int \frac{d^4p_E}{(2\pi)^4}
\frac{p_E\cdot(p+q)_E -\rho^2}{(p_E^2+\rho^2)((p+q)_E^2+\rho^2)}\right]
\right|_{q_E^2=0} \nonumber \\
&=& \frac{N_c}{4\pi^2}\left[\ln(\frac{\Lambda^2}{\rho^2}) -1\right] +
{\cal O}(\rho^2/\Lambda^2),
\label{e4}
\end{eqnarray}
while the potential $V$ reads
\begin{equation}
V(\rho^2) = N_c\left[-4 \int^{\Lambda^2}_{0}\frac{d^4p_E}{(2\pi)^4}
\ln(p_E^2+\rho^2) + \frac{\Lambda^2}{2g}\rho^2\right],
\label{e5}
\end{equation}
where $p_E, q_E$ denote the Wick rotated Euclidean momenta.
Throughout the article we compute as usual in the large $N_c$ limit.
When the coupling $g$ is sufficiently large the potential can have a
minimum at a nonvanishing $\rho^2=\rho_0^2$ and the symmetry is broken
in the large $N_c$ limit.
The minimum of the potential is determined by the gap equation
\begin{eqnarray}
\frac{1}{g}
&=& \frac{1}{2\pi^2}\int^{1}_{0} d t \frac{t}{t+\rho_0^2/\Lambda^2}
\nonumber \\
&=&  \frac{1}{2\pi^2}\left[1 -\frac{\rho_0^2}{\Lambda^2}
\ln(\frac{\Lambda^2+\rho_0^2}{
\rho_0^2})\right] 
\label{e6}
\end{eqnarray}
which has a nonvanishing $\rho_0^2$ solution provided $g>g_{{\rm cr}}$,
where $g_{{\rm cr}}=2\pi^2$.

\begin{figure} 
\begin{center}
\epsfig{file=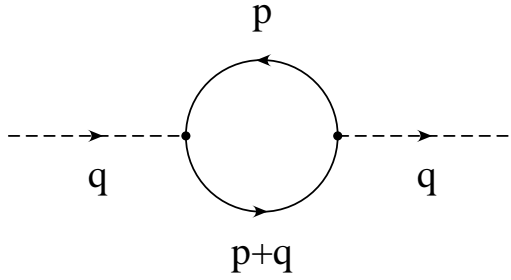, %
        height=4.0cm, angle=0}
\end{center}
\caption{Wave function normalization of the mesons by a quark-antiquark
loop.}
\end{figure}

Expanding about the vacuum $(\sigma,\pi_a)=(\rho_0,0)$, and putting
$\sigma=\sigma' +\rho_0$, we obtain the quadratic part of the Lagrangian
(the prime in $\sigma'$ is now omitted):
\begin{equation}
{\cal L} = \frac{ Z(\rho_0^2)}{2}\left[ (\partial_\mu\sigma)^2 -4 \rho_0^2
\sigma^2 +(\partial_\mu\pi_a)^2\right]+ \cdots
\end{equation}
which shows that $\sigma$ has a mass $m_{\sigma}=2 \rho_0$ and $\pi_a$
are the massless Nambu-Goldstone bosons.

KV observed that although the  potential (\ref{e5}) breaks the chiral
symmetry in the large $N_c$ limit, the quantum fluctuations of the mesons
can restore the symmetry when $N_c$ is smaller than a certain critical
value $N_c^{{\rm cr}}$.
This observation was made as follows.
The nonlinear sigma model of the massless modes associated with the
Lagrangian (\ref{e3}) is given by
\begin{equation}
{\cal L} = \frac{Z(\rho_0^2)}{2}\left[
\partial_\mu\Sigma^A\partial^{\mu}\Sigma^A
-\lambda (\Sigma^A\Sigma^A-\rho_0^2)\right],
\end{equation}
where $\Sigma^A\equiv(\sigma,\pi_a)$ and $\lambda(x)$ is the Lagrange
multiplier for the constraint $ \Sigma^A\Sigma^A-\rho_0^2=0$.
Integrating out the $\Sigma^A$ gives an effective function in $\lambda$
\begin{equation}
W(\lambda) = -2\int\frac{d^4p_E}{(2\pi)^4} \ln(p_E^2+\lambda)
+\frac{\lambda\beta}{2},
\label{e999}
\end{equation}
where $\beta\equiv Z(\rho_0^2)\rho_0^2$ is the stiffness.
Extremizing the function a second gap equation for $\lambda$ is obtained:
\begin{equation}
\frac{\Lambda^2}{8\pi^2}\left[\int^{1}_0 d\,t \frac{t}{t
+\lambda}\right]-\frac{\beta}{2}=0, 
\end{equation}
which has a nonvanishing solution $\lambda=\lambda_0$ when the stiffness
satisfies
\begin{equation}
\beta < \beta_{{\rm cr}},
\label{e12}
\end{equation}
where $\beta_{{\rm cr}}=\Lambda^2/4\pi^2$.
A nonvanishing solution for $\lambda$ implies that the $\Sigma^A$ obtain
a common mass $m_{\Sigma^A}^2=\lambda_0$ and, consequently, the chiral
symmetry remains unbroken.
Using (\ref{e4}) it can be checked that for $N_c =3$ the condition
(\ref{e12}) is satisfied when $\rho_0^2/\Lambda^2 <e^{-1}$, which implies
no chiral symmetry breaking in the real world case.
Here an important note is in order. 
In drawing the conclusion it was assumed that the UV cutoff in the mesonic
loop (in eq.(\ref{e999})), which is $1/N_c$ correction, was identical
 to that of the quark loops. Since the theory is nonrenormalizable 
there is, however, no reason that the assumption
should be true, and in fact the cutoffs should be assumed 
different \cite{dstl}. A recent study shows there is a possibility for the 
chiral symmetry restoration only when the cutoff
for meson loops is not small  compared to that of the quark
 loops \cite{obw99}.

{}From the discussion so far it is clear that the problem lies with too
small a wave function normalization $Z(\rho_0^2)$ for a given $\rho_0$.
We now suggest a modification of the NJL model in which the stiffness
can be sufficiently large so that $\beta>\beta_{{\rm cr}}$ even for $N_c=3$.
Back to (\ref{e2}), we see the easiest way to increase the wave function
normalization is to add
\begin{equation}
\frac{cN_c}{2} \left[(\partial_\mu \sigma)^2 +(\partial_\mu\pi_a)^2\right]
\label{e14}
\end{equation}
to the original Lagrangian, where $c$ is a positive constant.
This new term obviously shifts the  wave function normalization by the amount
\begin{equation}
\delta Z= c N_c.
\label{e100}
\end{equation}
Since the addition of (\ref{e14}) does not affect the potential
$V(\rho^2)$ the gap equation is not modified, and the shift (\ref{e100})
can render the stiffness greater than the critical value when $c$ is
sufficiently large.
Numerically one can obtain $N_c^{{\rm cr}} <3$ with $c=0.7/g$.

The addition of the term (\ref{e14}) corresponds to an addition of HDO's
to the original NJL Lagrangian (\ref{e1}).
To see this we may solve iteratively the equations of motion for the meson
fields arising from the Lagrangian (\ref{e2}) and (\ref{e14}), obtaining
\begin{eqnarray}
\sigma&=&-\frac{g}{N_c\Lambda^2} (1+c g \partial^2/\Lambda^2)^{-1}
\bar{\psi}\psi \nonumber \\ 
&=&-\frac{g}{N_c\Lambda^2} (1- c g \partial^2/\Lambda^2+\cdots
)\bar{\psi}\psi, \nonumber \\
\pi_a&=&-i\frac{g}{N_c\Lambda^2} (1+c g \partial^2/\Lambda^2)^{-1}
\bar{\psi}\gamma_5\tau_a\psi \nonumber \\ 
&=&-i\frac{g}{N_c\Lambda^2} (1- c g \partial^2/\Lambda^2+\cdots
)\bar{\psi}\gamma_5\tau_a\psi,
\end{eqnarray}
which leads to a  modified NJL Lagrangian:
\begin{eqnarray}
{\cal L}&=&\bar{\psi}i\partial\!\!\!/ \psi + \frac{g}{2N_c\Lambda^2}\left\{
\left[(\bar{\psi}\psi)^2 -(\bar{\psi}\gamma_5\tau_a\psi)^2\right] \right.
\nonumber \\
 & &\!\!\!\! - \left. \frac{cg}{\Lambda^2}\left[
(\bar{\psi}\psi) \partial^2( \bar{\psi}\psi)  -
(\bar{\psi}\gamma_5\tau_a\psi) \partial^2( \bar{\psi}\gamma_5\tau_a\psi)
\right]+\cdots
\right\}.
\end{eqnarray}
This clearly shows that the wave function normalization is sensitive to
the presence of HDO's.

We may now consider a more general modification of the NJL model:
\begin{equation}
{\cal L}= \bar{\psi}[ i\dslash - (\sigma
+i\gamma_5\tau_a\pi_a)K(\partial'^2/\Lambda^2,
\partial^2/\Lambda^2)]
\psi -\frac{N_c\Lambda^2}{2g}(\rho^2 +g\frac{{\cal O}_4}{\Lambda^2}),
\end{equation}
where
\begin{equation}
K(\partial'^2/\Lambda^2,\partial^2/\Lambda^2)=
1+ c_{ij}\left(\partial'^2/\Lambda^2\right)^i
\left(\partial^2/\Lambda^2\right)^j
\end{equation}
is a finite polynomial so that the Lagrangian is local, while
$c_{10}, c_{01},\cdots,$ are dimensionless constants.
Here $\partial'^2\equiv \stackrel{\leftarrow}{\partial_\mu}
\stackrel{\rightarrow}{\partial^\mu}$ and
\begin{equation}
{\cal O}_4=c_0 m^2 \rho^2  +
c_1 [(\partial_\mu\sigma)^2  +(\partial_\mu\pi_a)^2]
+c_2 \rho^4,
\end{equation}
where $c_i$ are dimensionless constants and the parameter $m$ is assumed
to be $m^2\! \sim \! \rho_0^2$.
The derivatives in $K$ are defined to act on the quark fields only.

Integration over the quark fields gives an effective Lagrangian
of the mesons in the form (\ref{e3}), now with  the wave function 
normalization and the potential given by
\begin{eqnarray}
Z(\rho^2)&=& -c_1N_c +\left. \frac{d}{dq_E^2}
\Sigma(q_E^2)\right|_{q_E^2=0}
\nonumber \\
&=& -c_1N_c +\left.\frac{d}{dq_E^2} \left[8 N_c \int 
\frac{d^4p_E}{(2\pi)^4}\frac{(p_E\cdot(p+q)_E
-\rho^2K_1K_2)K_1K_2}{(p_E^2+
\rho^2K_1^2)((p+q)_E^2+\rho^2K_2^2)}\right]
\right|_{q_E^2=0},
\label{e21}
\end{eqnarray}
where
\begin{eqnarray}
K_1&=&K(-p_E\cdot(p+q)_E/\Lambda^2,p_E^2/\Lambda^2), \nonumber \\
K_2&=&K(-p_E\cdot(p+q)_E/\Lambda^2,(p+q)_E^2/\Lambda^2),
\end{eqnarray}
and
\begin{equation}
V(\rho^2)= N_c\left[-4\int\frac{d^4p_E}{(2\pi)^4}
\ln[p_E^2+\rho^2K^2(-p_E^2/\Lambda^2,p_E^2/\Lambda^2)]
+\frac{\Lambda^2}{2g}\left(\rho^2+ g\frac{c_0m^2\rho^2
+c_2\rho^4}{\Lambda^2}\right)\right].
\label{e23}
\end{equation}

Minimization of the potential gives the gap equation as
\begin{eqnarray}
\frac{1}{g} \left(1+ g\frac{c_0 m^2
+ 2 c_2\rho_0^2}{\Lambda^2}\right)
&=& \frac{8}{\Lambda^2} \int \frac{d^4p_E}{(2\pi)^4}
\frac{K^2(-p_E^2/\Lambda^2,p_E^2/\Lambda^2)}
{p_E^2+\rho_0^2K^2(-p_E^2/\Lambda^2,p_E^2/\Lambda^2)
} \nonumber \\
&=& \frac{1}{2\pi^2}\int^{1}_{0} d t \frac{t
K^2(-t,t)}{t+\frac{\rho_0^2}{\Lambda^2} K^2(-t,t)}.
\label{e24}
\end{eqnarray}
Using this equation and (\ref{e6}) we can now obtain the shift in
the critical coupling due to the HDO's as
\begin{equation}
\delta(\frac{1}{g_{{\rm cr}}}) =h_g(c_{ij}) +{\cal O}(m^2/\Lambda^2),
\end{equation}
where
\begin{eqnarray}
h_g(c_{ij})&=&  \frac{1}{2\pi^2}\int^{1}_{0} d t \left[ K^2(-t,t)-1\right]
\end{eqnarray}
which shows that the critical coupling depends on the HDO's, unsuppressed
in $1/\Lambda^2$, as it should be since the integral in the gap equation
is quadratically divergent.

Now the wave function normalization in (\ref{e21}) can be written as 
\begin{equation}
Z(\rho^2)=Z_0(\rho^2) +h_z(c_{ij},c_1) +{\cal O}(\rho^2/\Lambda^2),
\label{e27}
\end{equation}
where $Z_0(\rho^2)$, the wave function normalization of the original NJL
model, is given by (\ref{e4}), and $h_z$ is a function depending only on
the coefficients of the HDO's.
This can be easily shown by noticing that the coefficient of the
logarithmically divergent term $\ln(\Lambda^2/\rho^2)$ in $Z(\rho^2)$ does
not depend on the presence of the HDO's, which can be seen by expanding
the integrand in (\ref{e21}) in powers of the coefficients $c_{ij}$ and
performing the integration.
This shows that the effect of HDO's in $K$ to the wave function
normalization is not suppressed by $1/\Lambda^2$ but ${\cal O}(1)$.

Expanding the potential (\ref{e23}) about the vacuum we obtain the
$\sigma$ mass
\begin{equation}
m_{\sigma}^2= \frac{V''(\rho_0^2)(2 \rho_0)^2}{Z(\rho_0^2)},
\end{equation}
where
\begin{eqnarray}
V''(\rho_0^2)&=&N_c\left[4
\int  \frac{d^4p_E}{(2\pi)^4} \frac{K^4(-p_E^2/\Lambda^2,p_E^2/\Lambda^2)}
{[p^2+\rho_0^2 K^2(-p_E^2/\Lambda^2,p_E^2/\Lambda^2)]^2} +
 c_2\right].
\label{e29}
\end{eqnarray}
As in the case of $Z(\rho^2)$, the coefficient of the logarithmically
divergent term in the integral is independent of the HDO's, and thus 
(\ref{e29}) may be written as
\begin{equation}
 V''(\rho_0^2)=Z_0(\rho_0^2)+ h_m(c_{ij},c_2) +{\cal O}(\rho_0^2/\Lambda^2),
\end{equation}
where $h_m(c_{ij},c_2)$ is a function of $c_{ij},c_2$ only.
Using (\ref{e27}),(\ref{e29}) we have
\begin{equation}
m_{\sigma}= 2 \rho_0 \sqrt{\frac{1+h_m/Z_0}{1+h_z/Z_0}}
\left[ 1+{\cal O}(\rho_0^2/\Lambda^2)\right].
\end{equation}
This shows that with the presence of HDO's the standard relation
$m_{\sigma}=2\rho_0$ is no longer valid \cite{suzuki90}.
Only when the coupling $g$ is fine tuned so that
$\rho_0^2/\Lambda^2 \ll 1$, $m_{\sigma}$ is about twice the
constituent quark mass.
In fact this is just one evidence for a more general conclusion.
The logarithmic divergence in (\ref{e21}), (\ref{e29}) and in the
coefficient of the marginal operator $\rho^4$ from the expansion of
the potential (\ref{e23}) shows that the effect of HDO's disappears
logarithmically in $\rho^2/\Lambda^2$ from the low energy effective
Lagrangian.
Of course this conclusion is derived from the large $N_c$ calculation,
and when $\rho_0^2/\Lambda^2 \ll 1$ the renormalization group running
of the parameters in the low energy effective Lagrangian due to the meson
fluctuations should be considered as in the top quark condensation
model \cite{bhl90,bardeen90,hill91,clv91}.
In the NJL model for QCD, however,  $\rho_0^2/\Lambda^2$ is not so small 
($ \rho_0 \sim 350$ MeV  and $\Lambda \sim 900$ MeV), and consequently
HDO's can play a role in determining the phase of the system.

In conclusion we have introduced HDO's in the NJL model through the
auxiliary meson fields and shown that they can have significant effects
on the low energy effective Lagrangian when the NJL coupling is such
that $\rho_0^2/\Lambda^2 \ll\!\!\!\!\!\!\!/ \,\,\, 1$.
In such case the low energy quantities such as the stiffness or
sigma meson mass 
are sensitive to the presence of HDO's and should be
regarded as free parameters unless sufficient informations are available
to determine the form and magnitude  of the HDO's.
This then allows one to avoid the chiral symmetry restoration suggested by
KV by modifying the NJL model with proper HDO's.

\vskip .3in
\noindent
{\bf Acknowledgments:} 
We are grateful to A. Grant, D. Hong, P. Ko, V.A. Miransky, C. Lee 
and W. Weise for useful 
discussions and especially wish to thank T. Clark for valuable comments.
This work was supported in part by the Korea Science and  Engineering
Foundation (KOSEF).


\end{document}